# Preliminary study on CAD-based method of characteristics for neutron transport calculation[*]

CHEN Zhen-Ping(陈珍平)[1;2] ZHENG Hua-Qing(郑华庆)[2] SUN Guang-Yao(孙光耀)[1;2]
SONG Jing(宋婧)[1;2] HAO Li-Juan(郝丽娟)[2] HU Li-Qin(胡丽琴)[1;2] WU Yi-Can(吴宜灿)[1;2]

[1]University of Science and Technology of China, Hefei of Anhui Prov., 230027, China
[2]Institute of Nuclear Energy Safety Technology, Chinese Academy of Sciences, Hefei of Anhui Prov., 230031, China

**Abstract:** The method of characteristics (MOC) is widely used for neutron transport calculation in recent decades. However, the key problem determining whether MOC can be applied in highly heterogeneous geometry is how to combine an effective geometry modeling method with it. Most of the existing MOC codes conventionally describe the geometry model just by lines and arcs with extensive input data. Thus they have difficulty in geometry modeling and ray tracing for complicated geometries. In this study, a new method making use of a CAD-based automatic modeling tool MCAM which is a CAD/Image-based Automatic Modeling Program for Neutronics and Radiation Transport developed by FDS Team in China was introduced for geometry modeling and ray tracing of particle transport to remove those limitations. The diamond-difference scheme was applied to MOC to reduce the spatial discretization errors of the flat flux approximation. Based on MCAM and MOC, a new MOC code was developed and integrated into SuperMC system, which is a Super Multi-function Computational system for neutronics and radiation simulation. The numerical results demonstrated the feasibility and effectiveness of the new method for neutron transport calculation in MOC.

**Key words:** method of characteristics; neutron transport calculation, CAD, geometry modeling

**PACS:** 28.20.Pr, 07.05.Tp

## 1 Introduction

Neutron transport calculation is one of the most important research areas in neutronics analysis of nuclear reactor design. With the rapid development of the nuclear reactor technology such as China Lead bismuth cooled Accelerator driven Reactor (CLEAR) and other Generation IV (GEN-IV) nuclear reactors, the requirements of nuclear computer codes for neutron transport calculation will be more and more challenged. Thus, the modern strategy for the analysis of advanced reactors must meet the following requirements: (a) ability to model multi-dimensional configuration with any degree of heterogeneity, (b) high accuracy and reasonable computing efficiency, (c) flexibility in energy group structure and cross-section processing, and (d) user-friendly interface and usability.

The method of characteristics (MOC) first proposed by Askew[1] has been considered as a potential candidate for meeting those challenged requirements mentioned above. Based on the integral-differential form of the neutron transport equation, the MOC combines the best advantages of the Collision Probability Method (CPM) and Discrete Ordinate Method (SN). Theoretically, it imposes no limitations on geometry configurations. Therefore, the MOC has already become one of the most important deterministic theories for neutron transport calculation with the rapid progress in computer science and technology. Therefore, many MOC codes were developed in the past twenty years such as CRX[2], CACTUS[3], CHAR[4], AutoMOC[5], etc. However, most computational algorithms based on MOC are geometry-dependent, which prevents their broader use to more heterogeneous calculations. The main problem is related to the geometry modeling associated with the ray tracing method. For instance, many codes were developed for particular geometry shapes and describe the geometry model with lines and arcs with a lengthy input data file, which imposes a number of limitations in further background meshing and ray tracing of the geometry domain. Therefore, the key determining whether the MOC can be applied in complicated and highly heterogeneous geometry is how to combine an effective geometry treatment method with MOC. In recent years, the solid modeling method with a great flexibility in description of the general geometry configurations is widely used for geometry modeling in MOC codes, such as ANEMOA[6] and AGENT[7]. In this study, under the framework of the CAD-based Multi-Functional 4D Neutronics Simulation System VisualBUS[8][9] developed by FDS Team, a new idea making use of the CAD/Image-based Automatic Modeling Program for

---

[*] Supported by the National Special Program for ITER (2011GB113006), the Strategic Priority Research Program of Chinese Academy of Sciences (XDA03040000) and the National Natural Science Foundation of China ( 91026004).
1) E-mail: zhenping.chen@fds.org.cn

Neutronics and Radiation Transport, which is named MCAM[10][11] developed by FDS Team, for geometry treatment was brought forward to solve the geometry problem mentioned above. Based on the theory and approach, a new MOC code was developed and integrated into the SuperMC system.

In this paper, the methodologies and numerical results for several benchmark problems will be presented. In section 2, the derivation of MOC equations from the general formalism of neutron transport equation is introduced briefly. Section 3 describes the geometry modeling method based on a powerful CAD modeling engine MCAM. The related ray tracing method is also described in this section. Section 4 presents the numerical results of several benchmark problems. Finally, the conclusions are summarized in section 5.

## 2 Method of characteristics

In the MOC, a huge amount of parallel straight lines will be implicitly produced on a system for certain discrete spatial directions as shown in Fig.1. These lines, known as characteristic lines, are regarded as neutron tracks along which the integral-differential formalism of the neutron transport equation reduces to the total derivative form. Before the derivation of the MOC transport equations, three basic assumptions are introduced: (a) the spatial domain is first partitioned into $N$ homogeneous regions in which the material properties are assumed to be constant, (b) the energy domain is divided into $G$ sub-energy groups and (c) the solid angle domain is subdivided into $M$ discrete directions with given discrete weights.

With those assumptions mentioned above, a ray tracing procedure is performed on this domain and generates a set of characteristic lines. The intersection of a characteristic line with the geometrical region will be referred as a trajectory. In Fig.1, each characteristic line represents a certain trajectory-based mesh in which the neutron flux is assumed to be flat distribution. Taking the one-group transport equation as an example, the neutron balance equation along the characteristic line can be written as:

$$\frac{d}{ds}\psi_{i,k}(s,\Omega_m) + \Sigma_{t,i}\psi_{i,k}(s,\Omega_m) = Q_i(\Omega_m) \quad (1)$$

Where $s$ is the distance away from the entering point; $\Sigma_{t,i}$ is the total macroscopic cross-section of the region $i$; $\psi_{i,k}(s,\Omega_m)$ is the angular flux in the region $i$ at distance $s$ along the $k$-th characteristic

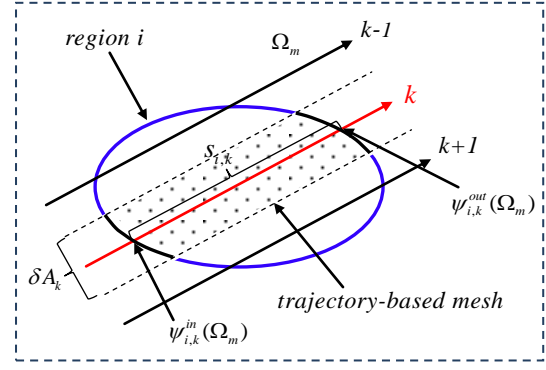

Fig. 1. The representation of MOC characteristic lines

line. $Q_i(\Omega_m)$ is the average neutron source in the region $i$.

The angular flux in the region $i$, $\psi_{i,k}(s,\Omega_m)$, along the line segment $k$ is then calculated by integrating Eq. (1) along the characteristic line.

$$\psi_{i,k}(s,\Omega_m) = \psi_{i,k}^{in}(\Omega_m)\exp(-\Sigma_{t,i}s) + \frac{Q_i(\Omega_m)}{\Sigma_{t,i}}\left(1-\exp(-\Sigma_{t,i}s)\right) \quad (2)$$

Where $\psi_{i,k}^{in}(\Omega_m)$ is the incoming angular flux at the entering point in the region $i$.

According to Eq. (2), the outcoming neutron angular flux from region $i$ along the line segment $k$ can be written as:

$$\psi_{i,k}^{out}(\Omega_m) = \psi_{i,k}^{in}(\Omega_m)\exp(-\Sigma_{t,i}s_{i,k}) + \frac{Q_i(\Omega_m)}{\Sigma_{t,i}}\left(1-\exp(-\Sigma_{t,i}s_{i,k})\right) \quad (3)$$

Where $s_{i,k}$ is the length of the $k$-th characteristic line.

Given the incoming angular flux $\psi_{i,k}^{in}(\Omega_m)$ and the outcoming angular flux $\psi_{i,k}^{out}(\Omega_m)$, by integrating the Eq. (2) along the $k$-th characteristic line from 0 to $s_{i,k}$, the segment average angular flux is obtained as:

$$\overline{\psi}_{i,k}(s,\Omega_m) = \frac{Q_i(\Omega_m)}{\Sigma_{t,i}} + \frac{\psi_{i,k}^{in}(\Omega_m) - \psi_{i,k}^{out}(\Omega_m)}{\Sigma_{t,i}s} \quad (4)$$

As shown in Fig.1, a single characteristic line represents one trajectory-based mesh where the neutron angular flux is assumed to be flat distribution. Therefore, the segment average angular flux, $\overline{\psi}_{i,k}(s,\Omega_m)$, is also the average angular flux of the trajectory-based mesh from the view point of the neutron balance equation.

Thus, given the average neutron angular flux of all the trajectory-based meshes in region *i*, the region average angular flux can be calculated with the Eq. (5):

$$\overline{\psi}_i(\Omega_m) = \frac{\sum_k \overline{\psi}_{i,k}(\Omega_m) s_{i,k} \delta A_k}{\sum_k s_{i,k} \delta A}  \quad (5)$$

Where $\delta A_k$ is the width of the segment *k* as shown in Fig.1.

Finally, the neutron scalar flux of the region *i* can be obtained as:

$$\phi_i = \sum_{m=1}^{M} \omega_m \overline{\psi}_i(s, \Omega_m) \quad (6)$$

Where $\omega_m$ is the weight for the direction $\Omega_m$, and *M* is the total number of the discrete directions.

## 3 Geometry modeling based on MCAM

The main problem limiting the broader usage of the most of MOC codes is associated with their ineffective and inefficient geometry treatment algorithm. In this section, a brief introduction about the geometry treatment based on MCAM will be presented.

### 3.1 Introduction to MCAM

MCAM[8][10][11][12] is a CAD/Image-based Automatic Modeling Program for Neutronics and Radiation Transport developed by FDS Team. It has been developed as an integrated interface program between commercial CAD systems and various radiation transport simulation codes, such as MCNP [13], TRIPOLI[14][15], GEANT4[16], FLUKA[17], and TORT[18]. On one hand, the engineering model created by CAD systems can be converted into the input geometry suitable for simulation codes conveniently. On the other hand, the exiting simulation model can be inverted into CAD model and visualized for further verification and updating. MCAM also supports a series of powerful supplementary functions such as creation and repair of CAD models and analysis of physics properties.

MCAM has already been applied to many complex nuclear facilities successfully, including the International Thermonuclear Experimental Reactor (ITER)[19], the super-conducting tokamak EAST[20] being operated in China, the FDS series reactors design[21] and Compact Reversed Shear Tokamak Reactor (CREST)[22], etc.

### 3.2 Geometry modeling

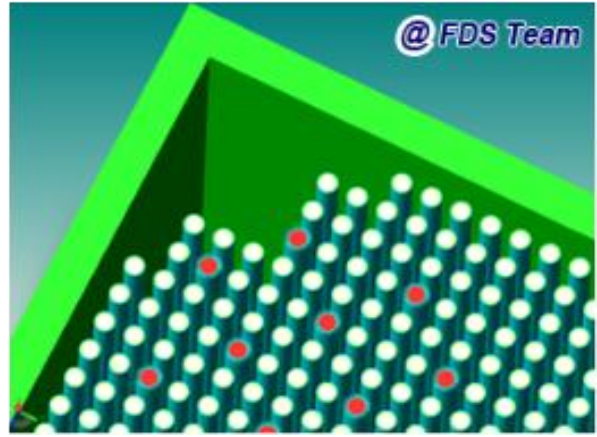

Fig. 2. The lattice geometry model visualized in MCAM

As discussed in section 3.1, one of the main functions of MCAM is its powerful geometry *creator* which supports the creation of various CAD geometry models. To construct the geometry model for MOC calculation, the Constructive Solid Geometry (CSG) which is a widely used method for solid modelers is introduced in MCAM for description of the geometry configuration. In MCAM, objects are built by copying, moving, slicing, rotating, arraying and mirroring of primitive objects such as cuboids, cylinders, spheres, cones and hexagonal prisms. Therefore, the configuration of the nuclear reactors can be constructed through Boolean operations (union, intersection, difference).

On one hand, the reactors geometry can be set up conveniently and rapidly with the geometry *creator* of MCAM. On the other hand, MCAM is compatible with the common intermediate formats (sat, step, igs) of CAD model which are supported by general commercial CAD systems such as CATIA, UG, and AutoCAD. In other words, the geometry model can be created by CATIA, UG, AutoCAD likewise MCAM. Moreover, the existing geometry model created by the commercial CAD modelers mentioned above can also be visualized in MCAM as shown in Fig.2 for further verification and updating.

### 3.3 Ray tracing algorithm based on MCAM

The ray tracing process is designed to generate the characteristic lines and obtain the related characteristic information of the geometry. In common sense, the ray tracers rely on mathematical solutions for the ray intersection an object and require different routines to be programed for various types of geometrical objects. Thus, the geometry-dependence of ray tracers was another key problem which prevents some MOC codes from broader use to some extent. With the rapid development of computer graphics, the first generalized ray tracer was developed in ANEMONA[6] which was based on the theory of R-functions. A generalized ray tracer which is geometry-independent does not need to recognize the specific geometry objects.

In this study, to remove the limitation mentioned above, a generalized ray tracer with C++ language based on the customization of ray-object intersection functions of MCAM was developed. The tracer performs the ray tracing process without hard coding for different geometrical entities. Theoretically, this kind of generality allows an arbitrary background meshing of the geometry model. In order to generate and get the characteristic lines information in SuperMC, the ray tracing process mainly includes the following steps: Firstly, import the CAD model being created by the modelers introduced in section 3.2 into MCAM for visualization and then perform the verification that whether the CAD model is coincident with the real geometry model. If necessary, updating is required for further repairing it. Secondly, background meshing for the CAD model is performed with irregular regions in which the material properties are assumed to be constant. Thirdly, the ray tracer searches for the ray-composite starting point walking along the ray direction from its starting point to ending point.

Then the intersection finding method will iteratively check whether the two consecutive intersection points along the ray are in the same region. If true, the segment between the two points is a valid characteristic line. Otherwise, the segment will be regarded as a virtual line which must be eliminated in ray tracing. Finally, after the foregoing three steps being implemented, the collector routine will collect the characteristic information such as the length of the characteristic line, region ID and material ID. These parameters are inevitably required for the subsequent MOC transport calculations.

## 4  Numerical validation

Based on the theories and methods as the foregoing statement, a new MOC code has been implemented in SuperMC. The numerical results for several problems will be given in the following.

### 4.1  ISSA problem

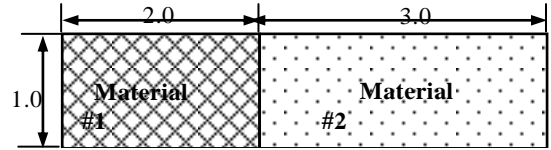

Fig. 3. The geometry configuration of ISSA problem (cm)

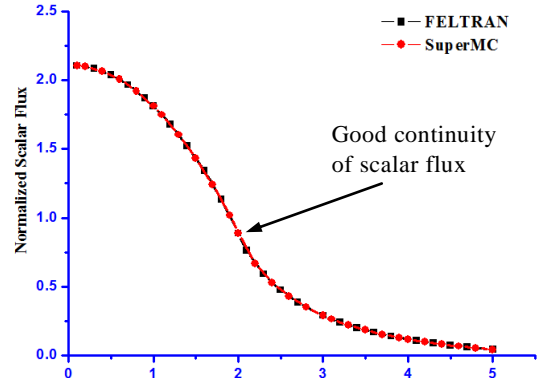

Fig. 4. The neutron flux distribution of ISSA problem

Table 1. Comparison of k-effective to the reference value for the ISSA problem

| Code | k-effective | Relative error (%) |
|---|---|---|
| ISSA | 1.67840 | reference value |
| FELTRAN | 1.67856 | 0.0095 |
| SuperMC | 1.67822 | - 0.0106 |

The first test case is the ISSA 1D problem[23] as shown in Fig.3. It just consists of two material regions whose geometry configurations are very simple. Supposing the geometry treatment as presented in section 3 is feasible and accurate for its simplicity of the geometry configurations, this problem was mainly used to verify SuperMC from the perspective of MOC itself. Only the right side of the problem has vacuum boundary condition while the other sides are reflective conditions. The macroscopic cross-sections of each material region are taken from the reference[24].

This problem was used for verifying the continuity of scalar flux and validity of the effective multiplication factor (k-effective) from the view point of MOC. The calculation results with SuperMC are compared with FELTRAN[24]. As seen in Fig.4, on the one hand, the computed scalar flux is in good agreement with the reference result given by FELTRAN. On the other hand, the computed scalar flux displays a smooth variation without any discontinuity at the interface between material #1 and material #2. Table 1 gives the comparison of eigenvalues with different codes to the reference value. The relative error of k-effective between SuperMC result and the reference value is about -0.0106%. In sum, both the neutron flux distribution and the k-effective are in good agreement with the reference results. It indicated that the MOC theory was accurately and effectively implemented in the code.

### 4.2 Multi-cell lattice rroblem

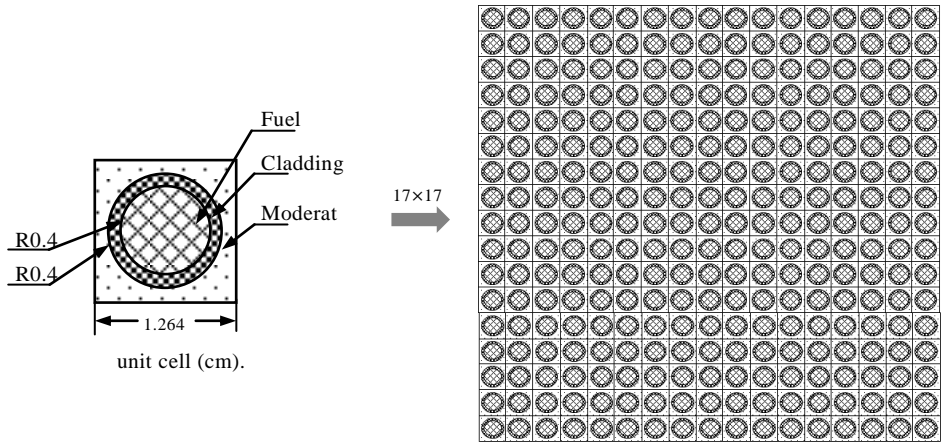

Fig. 5. The multi-cell lattice problem with its corresponding material compositions and geometry configurations

To verify the feasibility and validity of the geometry treatment of the code, a multi-cell lattice problem[5] like a PWR fuel assembly was used. Two levels calculation, i.e. the unit cell calculation and fully assembly calculation, were performed for the problem. As seen in Fig.5, the unit cell represents a small three region square cell having a side of 1.2647cm and consisting of two angular regions with outer radii of 0.41cm and 0.47cm, respectively. The inner region corresponds to a fissile material, the intermediate region to a cladding and the outer region to a moderating material. The macroscopic cross-sections of each material region are given in reference[5]. The multi-cell lattice geometry which has 17×17 pin cell arrangement is shown in the right side of Fig.5. All the pin cells are identical to the unit cell in geometry configuration and material composition.

Firstly, the k-infinite of unit cell was computed to verify the geometry treatment of irregular geometry configurations. Table 2 shows the numerical results with different codes for the unit cell. The calculated result from SuperMC is compared with that of CHAR-A[4] and TIBERE-2[25]. The relative error in k-infinite between SuperMC and CHAR is 0.0394% which is more accurate than that of TIBERE-2. Secondly, the multi-cell lattice geometry problem was also used for further verification of the code for treatment of large-scale geometries such as a full assembly. As seen in Table 3, the difference in k-infinite between SuperMC and MCNP is -0.0075%, which shows a better agreement with the reference value than that of AutoMOC. The parameter of k-infinite for the unit cell problem and the multi-cell lattice problem should be in a good agreement with each other from the prospective of neutron transport equation. Comparing the results between Table 2 and Table 3, although the k-infinite of the unit cell problem is larger than that of the multi-cell lattice problem, the two results still show a good agreement with each other corresponding the relative error about 0.0451%.

Table 2. Comparison of k-infinite with different codes for the unit cell geometry

| Code | k-infinite | Relative error (%) |
|---|---|---|
| CHAR-A | 1.06403 | reference value |
| TIBERE-2 | 1.06496 | 0.0874 |
| SuperMC | 1.06445 | 0.0394 |

Table 3. Comparison of k-infinite with different codes for the multi-cell lattice geometry

| Code | k-infinite | Relative error (%) |
|---|---|---|
| DORT | 1.06405 | reference value |
| AutoMOC | 1.06452 | 0.0442 |
| SuperMC | 1.06397 | - 0.0075 |

## 4.3 C5G7 MOX Benchmark

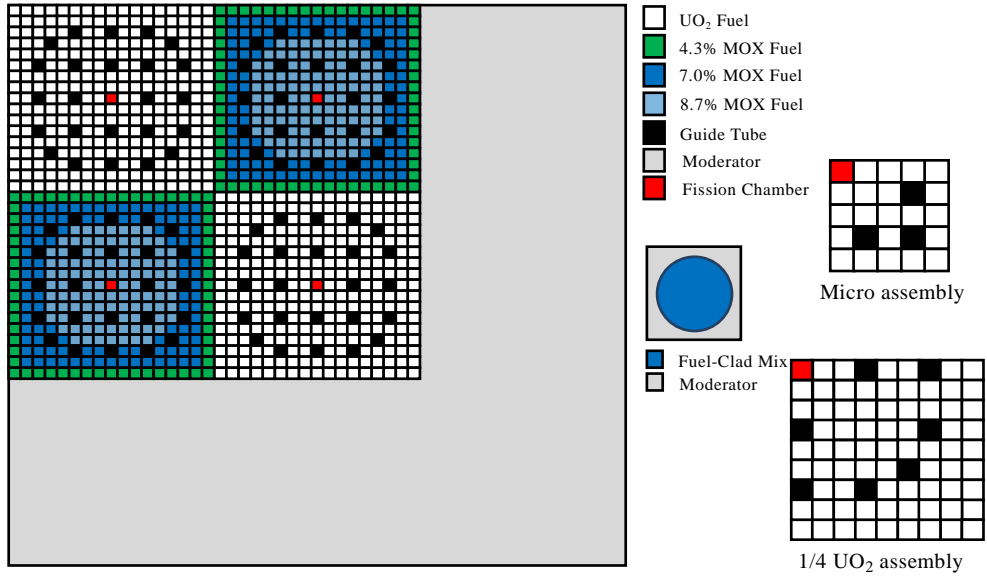

Fig. 6. The layout of the 2D C5G7 benchmark with UO2 and MOX assembly

Table4. Comparison of eigenvalues for C5G7 benchmark with its corresponding sub-models

| Code | Micro assembly | 1/4 $UO_2$ assembly | 1/1 $UO_2$ assembly | C5G7-MOX |
|---|---|---|---|---|
| GALAXY | — | 1.333776 | 1.333796 | 1.186660 |
| AGENT | 1.335200 | — | — | — |
| DeCART | 1.335060 | — | — | 1.186600 |
| SuperMC | 1.335033 | 1.331345 | 1.331782 | 1.182584 |

This benchmark[26] is a general problem to test the ability of modern deterministic methods and codes to treat such reactor core problems without spatial homogenization. The benchmark geometry chosen is the sixteen assembly (quarter core symmetry) C5 MOX fuel assembly problem specified by Cavarec[27]. It consists of two $UO_2$ fuel assemblies and two MOX fuel assembies. The two-dimensional (2D) configurations and material compositions are shown in Fig.6. As indicated, vacuum boundary conditions are applied to the right and to the bottom of the geometry while reflective boundary conditions are applied to the top and left of the geometry. The overall dimensions of the 2D configuration as shown are 64.26×64.26 cm, while each assembly is 21.42×21.42 cm. Each fuel assembly is made up of a 17×17 lattice of square pin cells, one of which is shown in Fig.6. The side length of each pin cell is 1.26 cm and all of the fuel pins and guide tubes have a 0.54 cm radius. As indicated by Fig.6, there are two compositions for every pin cell. A seven-group set of cross-sections was obtained from the literature[26].

To perform a more comprehensive validation, the C5G7 benchmark with its corresponding sub-models, i.e. micro assembly, 1/4 $UO_2$ assembly and 1/1 $UO_2$ assembly, were tested completely. The geometry configurations and material compositions of the several sub-models are illustrated in Fig.6. Table 4 shows the comparison of numerical results calculated with different codes. For the micro assembly, the result of SuperMC was compared with that of AGENT[7] and DeCART[28]. The maximum difference in k-infinite between SuperMC and other two codes is less than 12.5 pcm (1.0E-5). The computed results of 1/4 $UO_2$ assembly and 1/1 $UO_2$ assembly from SuperMC were compared with those of GALAXY[28]. The difference in k-infinite for the 1/4 $UO_2$ assembly and 1/1 $UO_2$ assembly are within 0.182% and 0.135%, respectively. The difference in SuperMC result between the 1/4 $UO_2$ assembly and 1/1 $UO_2$ assembly is about 32 pcm which shows a good agreement between the two models. The k-effective of the whole C5G7 MOX benchmark calculated by SuperMC was compared with that of GALAXY and DeCART. The reference value is 1.18655 given in the reference[26]. The difference in k-effective between the SuperMC result and reference value is about 0.33%. From the numerical analysis as the forgoing statement, the error will become bigger when the geometry configurations and material compositions of the model tend to be more complex and heterogeneous. The error may be mainly introduced by using a CAD model with MOC as opposed to the more traditional method of handling geometry and it will be fixed in the near future. Although the error becomes bigger, SuperMC still shows a reasonable correctness and accuracy when dealing with complex models.

* Supported by the National Special Program for ITER (2011GB113006), the Strategic Priority Research Program of Chinese Academy of Sciences (XDA03040000) and the National Natural Science Foundation of China ( 91026004).
1) E-mail: zhenping.chen@fds.org.cn

## 5 Conclusion

Under the framework of SuperMC, a new CAD-based MOC code for neutron transport calculation was developed. A detail description of the theoretical background and the generic computational algorithm used in the code was described. The methodology represents a unique synergistic combination of the method of characteristics and CAD technology. Therefore, thanks to the powerful capability of CAD modeling and ray tracing, the construction of complex geometry associated with ray tracing becomes quite efficient and convenient. The geometry can be constructed by general commercial CAD modeling tools (i.e. CATIA, UG, AutoCAD) besides MCAM. Thus a wide range of choices are available for users to choose a preferable modeling tool. At the same time, a geometry-independent ray tracer customized based on MCAM can perform the job without considering the specific geometry shapes, which indicates a great potential probability to apply the MOC to more complex models for transport calculations.

The numerical results show that the code has a reasonably good agreement with other codes which indicates that it can perform the neutron transport calculation correctly and automatically. In other words, the new method making use of MCAM for geometry treatment in MOC was prove to be of feasibility and effectiveness which indicated a broader usage of MOC to more complex models for neutron transport calculation in future.